\documentclass{llncs}

\usepackage[utf8]{inputenc}
\usepackage{graphicx}
\usepackage{amsmath}

\def\sys{\mbox{\textsc{LeoPARD}}}
\def\leoa{\mbox{\textsc{LEO-I}}}
\def\leob{\mbox{\textsc{LEO-II}}}
\def\leoc{\mbox{\textsc{Leo-III}}}

\title{\sys\ --- A Generic Platform for the Implementation of
 Higher-Order Reasoners\thanks{This work has been supported by
    the DFG under grant BE 2501/11-1 (\leoc). The final publication is available at \url{http://link.springer.com.}}}

\author{Max Wisniewski, Alexander Steen and Christoph Benzm\"uller}

\institute{Dept. of Mathematics and Computer Science, 
 Freie Universit\"at Berlin, Germany \\ \email{max.wisniewski|a.steen|c.benzmueller@fu-berlin.de}}

\begin{document}

\maketitle

\begin{abstract}
  \sys\ supports the implementation of knowledge
  representation and reasoning tools for higher-order logic(s).  It
  combines a sophisticated data structure layer (polymorphically typed $\lambda$-calculus with nameless spine
  notation, explicit substitutions, and perfect term sharing) with an
  ambitious multi-agent blackboard architecture (supporting
  prover parallelism at the term, clause, and search level). Further
  features of \sys\ include a parser for all TPTP
  dialects, a command line interpreter, and generic means for the
  integration of external reasoners.
\end{abstract}

 \section{Introduction}
 \sys\ (\textbf{L}eo's \textbf{P}arallel \textbf{AR}chitecture and
 \textbf{D}atastructures) is designed as a generic system
 platform for implementing higher-order  (HO) logic based knowledge
 representation, and reasoning tools. In particular, \sys\ provides
 the base layer of the new HO automated theorem prover
 (ATP) \leoc, the successor of the well known provers \leoa\ \cite{C3}
 and \leob\ \cite{C26}. 

Previous experiments with \leoa\ and the \textsc{OAnts} mechanism \cite{C8}
indicate 
a flexible, multi-agent blackboard
 architecture is well-suited for automating HO logic \cite{J16}. However, (due to
 project constraints) such an approach has not been realized in 
 \leob. Instead, the focus has been on the proof search layer in
 combination with a simple, sequential collaboration with an external
 first-order (FO) ATP.
 \leob\ also provides improved term data structures, term indexing,
 and term sharing mechanisms, which unfortunately have not
 been optimally exploited at the clause and the proof search
 layer.  For the development of \leoc\ the philosophy therefore has been to
 allocate sufficient resources for the initial development of a
 flexible and reusable system platform. The goal has been
 to bundle, improve, and extend the features with the highest potential of the
 predecessor systems \leoa, \leob\, and \textsc{OAnts}.

 \sloppy The result of this initiative is \sys \footnote{\sys\ can be download at:  
   \url{https://github.com/cbenzmueller/LeoPARD.git}.}, which is written in
 Scala and currently consists of approx. 13000 lines of
 code. 
 \sys\ combines a
 sophisticated data structure layer \cite{SteenMSc} (polymorphically typed $\lambda$-calculus with nameless spine
 notation, explicit substitutions, and perfect term sharing), with a
 multi-agent blackboard architecture \cite{WisniewskiMSc} (supporting prover parallelism at the term, clause, and search
 level) and  further tools including a parser for all TPTP 
 \cite{DBLP:journals/jar/Sutcliffe09,J22} syntax dialects, generic
 support for interfacing with external reasoners, and a command line
 interpreter.  Such a combination of features and support tools is, up to the
 authors knowledge, not matched in related HO reasoning frameworks.

 The intended users of the \sys\ package are implementors of HO
 knowledge representation and reasoning systems, including novel ATPs and
 model finders. In addition, we
 advocate the system as a platform for the integration and
 coordination of heterogeneous (external) reasoning tools.

\section{Term Data Structure}
Data structure choices are a critical part of a theorem prover and
permit reliable increases of overall performance when implemented and exploited
properly. Key aspects for efficient theorem proving have been an
intensive research topic and have reached maturity within FO-ATPs
\cite{riazanov2003implementing,Sekar:2001:TI:778522.778535}.
Naturally, one would expect an even higher impact
of the data structure choices in HO-ATPs. However, in the latter
context, comparably little effort has been invested yet --
probably also because of the inherently more complex nature of HO logic.

\paragraph{Term Language.} The \sys\ term language extends the simply typed $\lambda$-calculus
with parametric polymorphism, yielding the second-order
polymorphically typed $\lambda$-calculus (corresponding to $\lambda$2
in Barendregt's $\lambda$-cube~\cite{Barendregt91}). In particular, the system under
consideration was independently developed by
Reynolds~\cite{Reynolds74} and Girard~\cite{girard1972interpretation}
and is commonly called System F today.  Further extensions, for
example to admit dependent types, are future work.

Thus, \sys\ supports the following type and term language:\\[-.5em]
\begin{equation*}\begin{split}
  \tau, \nu  &::= t \in T                        \hspace{.5em} \text{\small(Base type)} \\[-.5em]
             & \quad | \; \alpha                 \hspace{2.5em} \text{\small(Type variable)} \\[-.4em]
             & \quad | \; \tau \to \nu           \hspace{.5em} \text{\small(Abstraction type)} \\[-.4em]
             & \quad | \; \forall \alpha. \; \tau \hspace{.8em}
             \text{\small(Polymorphic type)} \\[.5em]
  \end{split}\end{equation*}
\begin{equation*}\begin{split}
  s, t &::= X_\tau \in \mathcal{V}_\tau \hspace{1.65em} | \; c_\tau \in \Sigma \hspace{4.1em} \text{\small(Variable / Constant)} \\[-.4em]
       & \quad | \; (\lambda x_{\tau} \: s_\nu)_{\tau \to \nu}     
            \; | \; (s_{\tau \to \nu} \; t_\tau)_\nu  \hspace{2.4em} \text{\small(Term abstr. / appl.)} \\[-.4em]
       & \quad | \; (\Lambda \alpha \: s_\tau)_{\forall \alpha\;\tau} \hspace{.35em}
            \; | \; (s_{\forall \alpha\;\tau} \; \nu)_{\tau[\alpha/\nu]}
                                                      \hspace{.9em}
                                                      \text{\small(Type
                                                        abstr. / appl.)}  
  \end{split}\end{equation*}

\noindent An example term of this language is:\\
$$\Lambda \alpha\lambda P_{\alpha \to o}\;(( f_{\forall \beta \;(\beta \to o) \to o \to o} \; \alpha) \; (\lambda Y_\alpha\; P \; Y )) \; T_o.$$

\paragraph{Nameless Representation.} Internally, \sys\ employs a locally nameless representation (both at the type
and term level), that extends de-Bruijn indices to (bound) type
variables \cite{KRTU99}. The definition of de-Bruijn indices~\cite{Bruijn72} for type
variables is analogous to the one for term variables. Thus, the above example term is represented namelessly as
\[\left(\Lambda\lambda_{\underline{1} \to o}\;((f_{\forall(\underline{1} \to o) \to o \to o} \; \underline{1}) \; (\lambda_{\underline{1}}\; 2 \; 1 )) \; T_o\right)\]
where de-Bruijn indices for type variables are underlined.

\paragraph{Spine Notation and Explicit Substitutions.} On top of nameless terms,
\sys\ employs spine notation~\cite{CP03} and explicit substitutions
\cite{Explicit}. The first technique allows quick head symbol
queries, and efficient left-to-right traversal, e.g. for unification
algorithms. The latter augments the calculus with substitution closures that admit
efficient (partial) $\beta$-normalization runs. Internally, the above example reads
\[\Lambda \lambda_{\underline{1} \to o}\; f_{\forall(\underline{1} \to o)\to o \to o} \cdot (\underline{1};\lambda_{\underline{1}}\; 2 \cdot (1);T)\] where $\cdot$ combines function \emph{heads} to argument lists (\emph{spines}) in which
$;$ denotes concatenation of arguments.

\paragraph{Term Sharing/Indexing.} Terms are perfectly shared within
\sys, meaning that each term is only constructed once and then reused
between different occurrences. This does not only reduce memory
consumption in large knowledge bases, but also allows constant-time
term comparison for syntactic equality using the term's pointer to its
unique physical representation. For fast (sub-)term retrieval based on
syntactical criteria (e.g. head symbols, subterm occurrences, etc.)
from the term indexing mechanism, terms are kept in
$\beta$-normal $\eta$-long form.

\paragraph{Suite of Normalization Strategies.} \sys\ comes with a
number of different (heuristic) $\beta$-normalization strategies that
adjust the standard leftmost-outermost strategy with different
combinations of strict and lazy substitution composition
resp. normalization and closure construction. $\eta$-normalization is
invariant wrt.~$\beta$-normalization of spine terms and hence
$\eta$-normalization (to long form) is applied only once for each
freshly created term.


\paragraph{Evaluation and Findings.} A recent empirical evaluation
\cite{SteenMSc} has shown that there is \textit{no single best
  reduction strategy} for HO-ATPs. More precisely, for different TPTP
problem categories this study identified different best reduction
strategies. This motivates future work in which machine learning
techniques could be used to suggest suitable strategies.


\section{Multi-agent Blackboard Architecture}
In addition to supporting classical, sequential theorem proving
procedures, \sys\ offers means for breaking the global ATP
loop down into a set of subtasks that can be computed in
parallel. This also includes support for subprover parallelism as successfully
employed, for example, in Isabelle/HOL's Sledgehammer tool
\cite{Sledgehammer}. More generally, \sys\ is construed to enable
parallalism at various levels inside an ATP, including the term,
clause, and search level \cite{Bonacina01}. For this,  \sys\ provides a
flexible multi-agent blackboard architecture.




\paragraph{Blackboard Architecture.} 
Process communication in \sys\ is realized indirectly via a blackboard
architecture \cite{weiss:mas}. The \sys\ blackboard
\cite{WisniewskiMSc} is a collection of globally shared and accessible
data structures which any process, i.e. agent, can query and
manipulate at any time in parallel.  From the blackboard's perspective each process is
a specialist responsible for exactly one kind of problem. The blackboard is generic in the data structures, i.e. 
it allows the programmer to add various kinds data structures for any kind of data.
Insertion into the data structures is handled by the blackboard. Hence, each specialist can indeed by
specialized on a single data structure.

The \sys\ blackboard mechanism and associated data structures provide
specific support for nested and-or search trees, meaning that sets of formulae
can be split into (nested) and-or contexts. Moreover, for each
supercontext respective TPTP SZS status \cite{DBLP:journals/jar/Sutcliffe09}  information is automatically
inferred from the statuses of its subcontexts.

\paragraph{Agents.}
In \sys\ specialist processes can be modeled as agents  \cite{WisniewskiMSc}.
Classically, agents are composed of three
components: environment perception, decision making, and action
execution  \cite{weiss:mas}. 

The perception of \sys\ agents is trigger-based, meaning that each
agent is notified by a change in the blackboard. \sys\ agents are to
be seen as homomorphisms on the blackboard data together with a filter
when to apply an action. 
Depending on the perceived change of the resp. state of the blackboard an
agent decides on an action it wants to execute.

\paragraph{Auction Scheduler.} Action execution in \sys\ is
coordinated by an auction based scheduler, 
which implements an own approximation algorithm \cite{WisniewskiMSc}
for combinatorical auctions \cite{arrow}.
More precisely, each \sys\
agent computes and places a bid for the execution of its action(s).  The
auction based scheduler then tries to maximize the global benefit of the
particular set of actions to choose.  

This selection mechanism works uniformly for all agents that can be
implemented in \sys. Balancing the value of the actions is therefore
crucial for the performance and the termination of the overall system. 
A possible generic solution for the agents bidding is to apply machine
learning techniques  to optimize the bids for the best overall
performance. This is future work.

Note that the use of  advanced agent technology in
\sys\ is optional. A traditional ATP can still be
implemented, for example, as a single, sequential reasoner instantiating exactly one
agent in the \sys\ framework.


\paragraph{Agent Implementation Examples.}
For illustration purposes, some agent implementations have been
exemplarily included in the \sys\ package. For example, simple agents
for \emph{simplification}, \emph{skolemization}, \emph{prenex-form},
\emph{negation-normal-form} and \emph{paramodulation} are
provided. Moreover, the agent-based integration of external ATPs is
demonstrated and their parallelization is enabled by the \sys\ agent
framework. This includes agents embodying \leob\ and Satallax
\cite{Satallax} running remotely on the
SystemOnTPTP \cite{DBLP:journals/jar/Sutcliffe09} servers in
Miami. These example agents can be easily adapted for other TPTP
compliant ATPs.

Each example agent comes with an applicability filter,
an action definition and an auction value
computation.  
The provided agents suffice to illustrate the working principles of
the \sys\ multi-agent blackboard architecture to interested
implementors. After the official release of \leoc, further, more
sophisticated agents will be
included and offered for academic reuse.

\section{Other Components}
The \sys\ framework provides useful further components. For example, a
generic parser is provided that supports all TPTP syntax
dialects. Moreover, a command line interpreter supports fine grained
interaction with the system. This is useful not only for debugging but
also for training and demonstration purposes.  As pointed at above,
useful support is also provided for the integration of external
reasoners based on the TPTP infrastructure. This also includes
comprehensive support for the TPTP SZS result ontology. Moreover,
ongoing and future work aims at generic means for the transformation
and integration of (external) proof protocols, ideally by exploiting
results of projects such as ProofCert\footnote{See
  \url{https://team.inria.fr/parsifal/proofcert/}}.




\section{Related work}
There is comparably little related work to \sys, since higher-order theorem provers typically implement
their own data structures. Related systems  (mostly concerning term representation) include $\lambda$Prolog and Teyjus \cite{teyjus},
the Abella interactive theorem prover \cite{abella}, and the logical framework Twelf \cite{twelf}.

\paragraph{Acknowledgements.}
We thank the reviewers for their valuable feedback. Moreover, we thank Tomer Libal and the students of the \leoc~project
for their contributions to \sys.

\bibliographystyle{plain}

\end{document}